\documentclass[12pt]{article}
\usepackage[utf8]{inputenc}
\usepackage{tikz}

\usepackage{jheppub}

\title{Compactifying the Kerr-Schild Double Copy}
\author[a]{Ross Dempsey}
\author[b]{and Peter Weck}
\affiliation[a]{Joseph Henry Laboratories, Princeton University\\Princeton, NJ 08544, USA}
\affiliation[b]{Department of Physics and Astronomy, Johns Hopkins University\\
3400 North Charles Street, Baltimore, MD 21218, USA}

\emailAdd{sdempsey@princeton.edu}
\emailAdd{pweck1@jhu.edu}

\let\bar\overline

\begin{document}

\begin{abstract}
{
We show that the classical double copy relationship for Kerr-Schild spacetimes can be dimensionally reduced to give a natural notion of the double copy for Kaluza-Klein theory with gravity coupled to a gauge field and a dilaton.  Under dimensional reduction the Kerr-Schild (KS) ansatz becomes the stringy Kerr-Schild (sKS) ansatz introduced by Wu. This ansatz captures many Kaluza-Klein black hole solutions, including single-charge black holes arising in both gauged and ungauged supergravity theories. We identify the single copy gauge field and scalar field of an arbitrary sKS solution. We show that the boost-reduction procedure for generating charged black hole solutions can be generalized to any stationary KS seed, and used to combine the metric with the zeroth and single copies of that seed into a single sKS solution. Furthermore, we comment on multi-charge solutions that can be written in a multi-sKS form, proposing a double copy interpretation involving multiple single copy sectors.
}
\end{abstract}

\maketitle

\section{Introduction}

The double copy paradigm has long provided an intriguing perspective on the relationship between gauge theory and gravity. Extending at least as far back as the KLT relations between open and closed string amplitudes \cite{Kawai:1985xq}, it has recently generated enormous interest in the amplitudes program thanks to the BCJ double copy relating perturbative gravity amplitudes to amplitudes from gauge theories \cite{Bern:2008qj,Bern:2019prr,adamoSnowmassWhitePaper2022}. This striking and unexpected feature of perturbative gravity naturally leads to the question of whether similar correspondences exist for exact solutions.

The first answer to this question came from the Kerr-Schild ansatz, a metric of the form
\begin{equation}\label{eq:ks}
    g_{\mu\nu} = \eta_{\mu\nu} + \Phi k_\mu k_\nu,
\end{equation}
for a null geodesic vector field $k_\mu$. Many of the best-known exact solutions to general relativity, such as the Kerr metric or the Myers-Perry metric in higher dimensions, admit this form. It was shown in \cite{Monteiro:2014cda} that for any stationary vacuum metric of the form \eqref{eq:ks}, the single copy gauge field and zeroth copy scalar field
\begin{equation}
    A_\mu \equiv \Phi k_\mu \qquad \text{and}\qquad \Phi
\end{equation}
satisfy the source-free Maxwell equations and Klein-Gordon equation, respectively.

The Kerr-Schild double copy has since been applied and generalized in a wide variety of cases, including to arbitrary spacetime dimension, curved background metrics, or alternate settings such as Einstein-Maxwell theory, double field theory, or exceptional field theory \cite{Monteiro:2020plf,Chacon:2021hfe,Cho:2019ype,Bahjat-Abbas:2020cyb,Berman:2020xvs,Luna:2015paa,Carrillo-Gonzalez:2017iyj,CarrilloGonzalez:2019gof,Luna:2016due,Bahjat-Abbas:2017htu,Alkac:2021bav,Berman:2018hwd,Easson:2022zoh,Angus:2021zhy}. Similar classical double copy relations have also been found for any algebraically special spacetime of Type D, by formulating the map in terms of the Weyl tensor and field strength tensors instead of the fields \cite{Luna:2018dpt}. This formulation and its twistorial generalization \cite{White:2020sfn} have led to better understanding of how the double copy relates solutions in the multipole expansion \cite{Chacon:2021hfe}, and why the Weyl or Kerr-Schild double copies can be written as a local map in position space \cite{Luna:2022dxo}.

Here we elaborate on another way to generalize the Kerr-Schild double copy, via dimensional reduction. We show in Section \ref{sec:main} how a Kaluza-Klein reduction of the Kerr-Schild ansatz in $D+1$ dimensions can be written as the $D$-dimensional stringy Kerr-Schild (sKS) ansatz introduced by \cite{Wu:2011zzh,Wu:2015nra}. This metric ansatz captures all known black hole solutions in ungauged supergravity theories, and many charged, rotating black hole solutions in gauged supergravity theories as well. We show that by using the same dimensional reduction on the single and zeroth copies of the $D+1$-dimensional Kerr-Schild solution, we obtain a natural notion of the single and zeroth copies of the sKS solution. This procedure extends the double copy to any sKS solution of Kaluza-Klein gravity on a flat background, and moreover we show that a similar prescription holds on a maximally symmetric background. In Section \ref{sec:generating}, we study some examples of the sKS double copy, and show how they can be generated from Kerr-Schild solutions using the boost-reduction procedure of \cite{frolovChargedRotatingBlack1987a}. We also comment on multi-charge solutions, whose behavior demonstrates that the sKS double copy can be extended to a class of AdS multi-charge black holes.

\section{Stringy Kerr-Schild and the Double Copy}\label{sec:main}

In Section~\ref{sec:reduction}, we show how a Kaluza-Klein (KK) reduction of a Kerr-Schild spacetime gives rise to the ``stringy Kerr-Schild'' (sKS) ansatz introduced in \cite{Wu:2011zzh} for Einstein-Maxwell-dilaton theory. 
We give the explicit rules for the sKS double copy in Section~\ref{sec:sks_double_copy}, and present an example for the black hole metric appearing in the 5D compactification of the D1/D5 system in Section~\ref{sec:d1d5}. Finally, in Section~\ref{sec:ads}, we show that an analogous correspondence follows on an AdS background spacetime, provided we give a mass to the zeroth copy scalar.

\subsection{$S^1$ Reduction of Kerr-Schild Spacetimes}\label{sec:reduction}

We begin by fixing conventions for the dimensional reduction of a pure gravity theory in $D+1$ dimensions to an Einstein-Maxwell-dilaton theory in $D$ dimensions. Let Latin indices run $a = 0,\ldots,D$ and Greek indices run $\mu = 0,\ldots,D-1$. We take the $(D+1)$-dimensional spacetime to have topology $\mathbb{R}^{1,D} \times S^1$, where the $S^1$ is an isometry and aligned with the $x^D$ direction. We then write its metric in the form
\begin{equation}\label{eq:kk}
    g^{(D+1)}_{ab}\,dx^a\,dx^b = e^{-(D-2)\phi}\left(dx^D + A_\mu\,dx^\mu \right)^2 + e^\phi g^{(D)}_{\mu\nu}\,dx^\mu\,dx^\nu.
\end{equation}
The relative power of $e^{\phi}$ is chosen so that if the higher-dimensional theory has the action
\begin{equation}
    S_{D+1} = \int d^{D+1}x\,\sqrt{-g^{(D+1)}} R^{(D+1)},
\end{equation}
then substituting this ansatz gives the action for the reduced theory in an Einstein frame\footnote{Here and throughout we ignore volume factors coming from the compactification, which are irrelevant for the classical equations of motion.}:
\begin{equation}\label{eq:emd_action}
    S_D = \int d^{D}x\,\sqrt{-g^{(D)}}\left\lbrack R^{(D)} - \frac{1}{4} e^{-(D-1)\phi} F_{\mu\nu}F^{\mu\nu} - \frac{(D-1)(D-2)}{4}\left(\nabla \phi\right)^2\right\rbrack.
\end{equation}

Now we apply this reduction procedure to a Kerr-Schild spacetime of the form \eqref{eq:ks}, as exhibited in \cite{Ett:2015fhw}. We take $\phi$ and $k_a$ to be independent of $x^D$, and furthermore for convenience we fix the normalization of $\phi$ so that $k_D = 1$. We then have
\begin{equation}\label{eq:ks_reduction}
\begin{split}
    g^{(D+1)}_{ab}\,dx^a\,dx^b &= \left(\eta_{ab} + \Phi k_a k_b\right)dx^a\,dx^b\\
    &= \left(1 + \Phi\right)\left(dx^D + \frac{\Phi}{1 + \Phi}k_\mu\,dx^\mu\right)^2 + \left(\eta_{\mu\nu} + \frac{\Phi}{1 + \Phi}k_\mu k_\nu\right)\,dx^\mu dx^\nu.
\end{split}
\end{equation}
The dimensionally-reduced Einstein-Maxwell-dilaton solution is then seen to be
\begin{equation}\label{eq:sks_reduced}
    \begin{split}
        g^{(D)}_{\mu\nu} &= e^{-\phi}\left[\eta_{\mu\nu} + \left(1-e^{(D-2)\phi} \right)k_\mu k_\nu\right],\\
        A_\mu &= \left(1-e^{(D-2)\phi}\right)k_\mu,\\
        e^{-(D-2)\phi} &= 1 + \Phi.
    \end{split}
\end{equation}
By construction the reduced vector $k_\mu$ must be timelike and satisfy $k_\mu k^\mu = -1$.

Note that the closely related compactification of a Kerr-Schild solution along a timelike circle was considered previously in \cite{Banerjee:2019saj}. In addition, the compactification of a Kerr-Schild solution on a spacelike torus has been studied in the context of half-maximal ungauged supergravity theories using double field theory, and the reduced theory was found to admit a Maxwell-scalar single copy description \cite{Angus:2021zhy}. A special feature of our setup is that the solution \eqref{eq:sks_reduced} matches the stringy Kerr-Schild (sKS) ansatz introduced in \cite{Wu:2011zzh}. This ansatz is known to encompass a wide range of solutions arising in KK supergravity theories, included gauged supergravity \cite{Wu:2015nra} (see Section \ref{sec:ads}). We will discuss some examples in Sections \ref{sec:d1d5} and \ref{sec:generating}.

\subsection{sKS Double Copy}\label{sec:sks_double_copy}

In $D+1$ dimensions, we know that the metric \eqref{eq:ks_reduction} has a single copy gauge field
\begin{equation}\label{eq:upper_gauge}
    \mathcal{A}^{(D+1)}_a = \Phi k_a.
\end{equation}
To construct a notion of the single copy of an sKS spacetime, we should reduce this gauge field in the same way we reduced the spacetime \eqref{eq:ks_reduction}. The appropriate KK ansatz in this case is simply
\begin{equation}\label{eq:vector_reduction}
    \mathcal{A}^{(D+1)}_a\,dx^a = \Phi\,dx^D + \mathcal{A}_\mu\,dx^\mu,
\end{equation}
which takes a gauge theory in $D+1$ dimensions to a gauge theory with an uncoupled scalar in $D$ dimensions:
\begin{equation}\label{eq:md_action}
    S_{D+1} = \int d^{D+1}x\,\left(-\frac{1}{4}\mathcal{F}_{ab}\mathcal{F}^{ab}\right) \quad \mapsto \quad S_D = \int d^D x\,\left(-\frac{1}{4}\mathcal{F}_{\mu\nu}\mathcal{F}^{\mu\nu} - \frac{1}{2}\left(\nabla \Phi\right)^2\right).
\end{equation}
When applied to \eqref{eq:upper_gauge}, we find a gauge field $\mathcal{A}_\mu = \Phi k_\mu$ and a scalar $\Phi$.

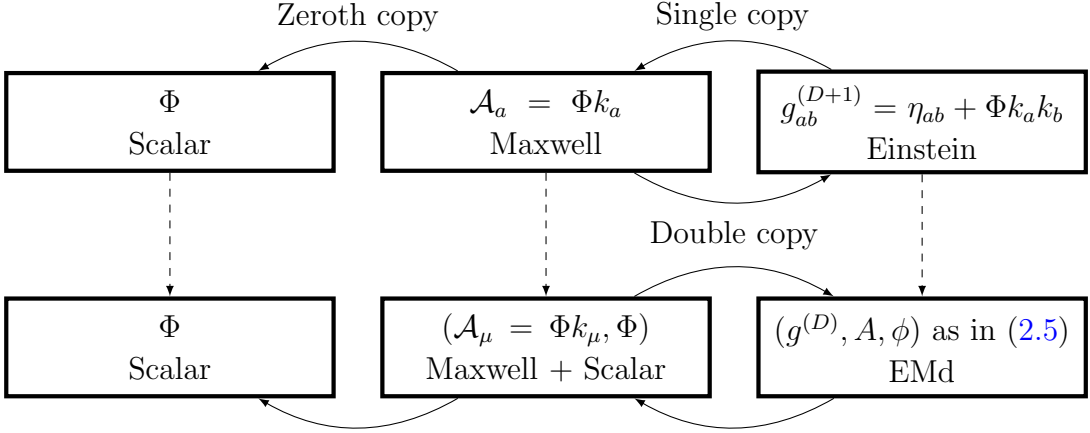
\begin{figure}
    \centering
    \begin{tikzpicture}[lower/.style={draw,ultra thick,rectangle,inner sep=5,black,text width=4cm,align=center,minimum height=1.3cm},upper/.style={draw,ultra thick,rectangle,inner sep=5,black,text width=4cm,align=center,minimum height=1.3cm}]
        \node[lower] at (0,0) (a) {$\Phi$\\Scalar};
        \node[lower] at (5,0) (b) {$(\mathcal{A}_\mu = \Phi k_\mu, \Phi)$\\Maxwell + Scalar};
        \node[lower] at (10,0) (c) {$(g^{(D)}, A, \phi)$ as in \eqref{eq:sks_reduced}\\EMd};
        \node[upper] at (0,3) (d) {$\Phi$\\Scalar};
        \node[upper] at (5,3) (e) {$\mathcal{A}_a = \Phi k_a$\\Maxwell};
        \node[upper] at (10,3) (f) {$g^{(D+1)}_{ab} = \eta_{ab} + \Phi k_a k_b$\\Einstein};
        
        \draw[dashed,-latex] (d) -- (a);
        \draw[dashed,-latex] (e) -- (b);
        \draw[dashed,-latex] (f) -- (c);
        
        \draw[-latex] (b) to[bend left=30] (a);
        \draw[-latex] (b) to[bend left=30] (c);
        \draw[-latex] (c) to[bend left=30] (b);
        
        \draw[-latex] (e) to[bend right=30] node[above] {Zeroth copy} (d);
        \draw[-latex] (e) to[bend right=30] (f);
        \node at (7.5, 1.5) {Double copy};
        \draw[-latex] (f) to[bend right=30] node[above] {Single copy} (e);
    \end{tikzpicture}
    \caption{Illustration of the chain of double copy relations between the various theories in question. Solid arrows represent double or single copy maps, and dotted arrows represent Kaluza-Klein reductions on the $S^1$.}
    \label{fig:theories}
\end{figure}

Comparing this with \eqref{eq:sks_reduced}, we see that we can identify the fields
\begin{equation}\label{eq:sks_sc}
    \mathcal{A}_\mu \equiv e^{-(D-2)\phi} A_\mu, \qquad \Phi \equiv e^{-(D-2)\phi} - 1
\end{equation}
as the single copy of an Einstein-Maxwell-dilaton solution of sKS form. Likewise, reducing the zeroth copy $\Phi$ from $D+1$ dimensions to $D$ dimensions, we find that $\Phi$ itself (defined as in \eqref{eq:sks_sc}) should be identified as the zeroth copy of an sKS solution.

Figure \ref{fig:theories} shows the various theories and field content we are considering. Starting at the upper right with a Kerr-Schild solution to pure gravity in $D+1$ dimensions, we are guaranteed to have a single copy solution to pure Maxwell theory and a zeroth copy solution to the Klein-Gordon equation, all in $D + 1$ dimensions. After compactifying the $x^D$ direction, we are then guaranteed solutions of the dimensionally reduced theories: the Einstein-Maxwell-dilaton theory \eqref{eq:emd_action}, a gauge theory with an uncoupled scalar in $D$ dimensions, and a scalar in $D$ dimensions, respectively. 

In summary, solely from the equations of motion of pure gravity in $D+1$ dimensions for the Kerr-Schild solution we start with, the fields \eqref{eq:sks_reduced} are guaranteed to solve \eqref{eq:emd_action} and the fields \eqref{eq:sks_sc} likewise solve the right-hand side of \eqref{eq:md_action}. Conversely, any sKS solution of the form \eqref{eq:sks_reduced} can be expressed as the dimensional reduction of a Kerr-Schild spacetime in $D+1$ dimensions. It follows that for any sKS solution \eqref{eq:sks_reduced}, the single copy \eqref{eq:sks_sc} solves the source-free Maxwell and Klein-Gordon equations.


Nevertheless, it is instructive to verify directly that the equations of motion for the Einstein-Maxwell-dilaton solution \eqref{eq:sks_reduced} imply those for the fields \eqref{eq:sks_sc}. Variation of the action \eqref{eq:emd_action} with respect to $(g^{\mu \nu}, A_\mu, \phi)$ yields equations of motion
\begin{align} \label{sKS_eqom}
&R_{\mu \nu}-\frac{(D-1)(D-2)}{4} \nabla_\mu \phi \nabla_\nu \phi -\frac{1}{2} e^{-(D-1)\phi} \left[F_{\mu \sigma}F_{\nu}^\sigma-\frac{1}{2(D-2)}g_{\mu \nu} F^2 \right] =0
\\ 
& \nabla^\mu\left[e^{-(D-1)\phi} F_{\mu \nu} \right]=0,  \qquad \Box \phi+\frac{e^{-(D-1)\phi}}{2(D-2)}  F^2=0 \nonumber.
\end{align}
Applying the method of background expansion outlined in \cite{Wu:2015nra}, these can be written in terms of the background metric and background covariant derivatives. As in \cite{Wu:2015nra}, we find that together with the geodesic condition $k^\mu \bar{\nabla}_\mu k^\nu=0$ the equations of motion \eqref{sKS_eqom} are equivalent to
\begin{equation}
s=0, \quad v_\mu=0, \quad t_{\mu \nu}=0,
\end{equation}
where we have introduced scalar, vector, and symmetric tensor objects defined with respect to the background spacetime (in this case $\eta_{\mu \nu}$)
\begin{equation}\label{eq:sKS_eqom_reduced}
\begin{split}
s &= \bar{\Box} \phi-(D-2) (\bar{\nabla} \phi)^2
\\
v_\mu &= \bar{\nabla}^\nu \left[2(e^{-(D-2)\phi}-1) \bar{\nabla}_{[\nu} k_{\mu]}+(D-2)e^{-(D-2)\phi}k_\nu \bar{\nabla}_\mu \phi\right] \\
&\quad -(D-2) e^{-(D-2)\phi} \bar{\nabla}^\nu \phi \bar{\nabla}_\nu k_\mu
\\
t_{\mu \nu} &= \bar{\nabla}^\lambda \left[(1-e^{-(D-2)\phi} )k_\lambda \bar{\nabla}_{(\mu}k_{\nu)} \right]-2(1-e^{-(D-2)\phi} )\bar{\nabla}_{[\lambda}k_{(\mu ]} \bar{\nabla}^\lambda k_{\nu)}.
\end{split}
\end{equation}
As pointed out in \cite{Wu:2015nra}, it is striking that the sKS dilaton and vector are governed by relatively simple equations with respect to the background metric. The double copy gives a new perspective on why this is the case. Using the single copy identifications \eqref{eq:sks_sc}, we can write
\begin{equation}
s=-\frac{1}{(D-2)} \frac{\bar{\Box}\Phi}{1+\Phi}, \qquad v_\mu=-\bar{\nabla}^\nu \mathcal{F}_{\mu \nu}-k_\mu \bar{\Box}\Phi.
\end{equation}
Clearly $s=0$ and $v_\mu=0$ if and only if $\bar{\Box}\Phi=0$ and $\bar{\nabla}^\nu \mathcal{F}_{\mu \nu}=0$. In other words, the source-free equations of motion governing the sKS ansatz \eqref{eq:sks_reduced} imply the source-free Maxwell and Klein-Gordon equations for the single and zeroth copy fields $(\mathcal{A}_\mu, \Phi)$, just as expected.

\subsection{Single Copy of the D1/D5 Black Hole}\label{sec:d1d5}

In the context of black hole microstate counting, a famous example of a black hole solution is that associated to the D1/D5 system of type IIB string theory \cite{Mandal:2000rp,callan:1996}. The string theory is compactified on a five-torus $T^5$, with $Q_5$ D5-branes wrapped on the $T^5$ and another $Q_1$ D1-branes wrapped on an $S^1$ embedded in the $T^5$. The branes are also given $N$ units of momentum around this $S^1$. As an example of the sKS double copy relations, we will give the single copy of the extremal black hole metric which can be obtained from the reduction of this system to five dimensions. More involved examples are treated in Section \ref{sec:generating}.

In the supergravity limit in the 5D compactification of this 10D type IIB configuration, we have a metric of the form
\begin{equation}\label{eq:d1d5_metric}
    ds^2 = -\lambda^{-2/3}\,dt^2 + \lambda^{1/3}\left(dr^2 + r^2 \,d\Omega_3^2\right),
\end{equation}
where
\begin{equation}
    \lambda = \prod_{i=1}^3 \left\lbrack 1 + \left(\frac{r_i}{r}\right)^2\right\rbrack
\end{equation}
and the three parameters $r_1$, $r_2$, and $r_3$ can be expressed in terms of the parameters $Q_1$, $Q_5$, and $N$ \cite{callan:1996}.

If we set all but one of these charges to zero\footnote{Note that this is a singular limit, in which the horizon area and the Bekenstein-Hawking entropy of the black hole vanish. In Section \ref{sec:multicharge}, we explore the notion of a single copy for solutions containing multiple nonzero charges.}, say $r_2 = r_3 = 0$, then we can write this metric as part of an sKS solution to the action \eqref{eq:emd_action}. If we take $r_1 = \mu \sinh^2\beta$ and set $e^{-3\phi} = \lambda = 1 + \frac{\mu \sinh^2\beta}{r^2}$, then the metric \eqref{eq:d1d5_metric} can be written in sKS form as
\begin{equation}
    ds^2 = e^{-\phi}\left[-dt^2 + dr^2 + r^2\,d\Omega_3^2 + \left(1 - e^{3\phi}\right)\left(\frac{\cosh\beta\,dt + dr}{\sinh\beta}\right)^2\right].
\end{equation}
Likewise, the gauge field is
\begin{equation}
    A_\mu = \left(1 - e^{3\phi}\right) k_\mu \quad\text{with}\quad k_\mu dx^\mu = \frac{\cosh\beta\,dt + dr}{\sinh\beta}.
\end{equation}

It follows from \eqref{eq:sks_sc} that the single copy fields are $\Phi = e^{-3\phi} - 1$, which solves the Klein-Gordon equation on a flat 5D spacetime, and the gauge field
\begin{equation}
    \mathcal{A}_\mu\,dx^\mu = \left(e^{-3\phi} - 1\right)k_\mu\,dx^\mu = \frac{\mu \sinh\beta}{r^2}\left(\cosh\beta \,dt + dr\right).
\end{equation}
After a gauge transformation, we see that this is the Coulomb field of a charge $\mathcal{Q} = \mu \sinh\beta \cosh\beta$.

\subsection{(A)dS Backgrounds}\label{sec:ads}

The sKS ansatz also encompasses many solutions on maximally symmetric backgrounds. For example, the ansatz was used in \cite{Wu:2011zzh} to construct a general family of rotating, single-charge AdS black holes with arbitary angular momenta. The uncharged, rotating solutions found in \cite{Gibbons:2004} and the single-charge limits of the static black hole solutions in \cite{cvetic:1999} can be obtained as special cases. 

By following the same chain of relationships illustrated in figure \ref{fig:theories}, we can understand how to write the single and zeroth copies of an sKS solution on an (A)dS background. The same single and zeroth copy maps apply. As we will demonstrate, however, the zeroth copy scalar $\Phi$ picks up a mass proportional to the curvature of the background.

Consider a $D$-dimensional anti-de Sitter background. The sKS ansatz in question is given by \eqref{eq:sks_reduced} with the Minkowski metric $\eta_{\mu \nu}$ replaced by a pure AdS$_D$ metric. The associated Einstein-Maxwell-dilaton action differs from \eqref{eq:emd_action} by a term coupling the dilaton to the cosmological constant, 
\begin{equation}
\begin{split}   
    S_D = \int d^{D}x\,\sqrt{-g^{(D)}}\Big\lbrace &R^{(D)} - \frac{1}{4} e^{-(D-1)\phi} F_{\mu\nu}F^{\mu\nu} - \frac{(D-1)(D-2)}{4}\left(\nabla \phi\right)^2
    \\
    &+g^2(D-1)e^\phi\left[D-3+e^{-(D-2)\phi}\right]\Big\rbrace,
\end{split}
\end{equation}
with $g$ the inverse AdS$_D$ radius. If we use the reduction ansatz \eqref{eq:ks_reduction} to uplift to a $(D+1)$-dimensional Kerr-Schild metric, the resulting spacetime is neither vacuum nor constant curvature. Nevertheless, we can follow the usual prescription to construct its single copy. If we assume the sKS solution and therefore its uplift are stationary, we can use the methods of \cite{Carrillo-Gonzalez:2017iyj} or \cite{Bah:2019sda} to determine the source of this gauge field using the timelike Killing vector of the metric. We find a current source
\begin{equation}
j^\mu=0, \quad j^{D}=-2g^2(D-3) \Phi.
\end{equation}
Note that the AdS$_D \times S^1$ background of the $(D+1)$-dimensional spacetime requires stress-energy content in addition to the $D$-dimensional cosmological constant, which under the double copy corresponds to a current along the compact $S^1$.

In the reduction to $D$ dimensions on the $S^1$, the $a=0,\cdots, D-1$ components of the $(D+1)$-dimensional Maxwell equation become the equation for a source-free $D$-dimensional Maxwell field, while the $a=D$ component can be interpreted as the Klein-Gordon equation for a scalar of mass $m^2 = 2g^2(D-3)$:
\begin{equation}
\bar{\nabla}_b \mathcal{F}^{ab}=-j^a \quad \Rightarrow \quad \bar{\nabla}_\nu \mathcal{F}^{\mu \nu}=0, \quad \bar{\Box} \Phi = -2g^2(D-3) \Phi,
\end{equation}
where the covariant derivatives on the right-hand side are with respect to the AdS$_D$ background. This is the same mass as given in \cite{Carrillo-Gonzalez:2017iyj} for the zeroth copy scalar of the ordinary Kerr-Schild double copy on an AdS$_D$ background.

The equations of motion satisfied by the sKS single and zeroth copy can also be inferred using the background expansion method of \cite{Wu:2015nra} for the sKS ansatz. The scalar condition in \eqref{eq:sKS_eqom_reduced} governing the sKS dilaton picks up an additional contribution, corresponding to the expected mass term for $\Phi$:
\begin{equation}
\begin{split}
s &= \bar{\Box} \phi-(D-2) (\bar{\nabla} \phi)^2-2g^2 \left(\frac{D-3}{D-2}\right)(1-e^{(D-2)\phi})  \\
&=-\frac{(1+\Phi)^{-1}}{(D-2)} \left(\bar{\Box}\Phi+2g^2(D-3)\Phi \right).
\end{split}
\end{equation}
As before, a source-free Maxwell equation for $\mathcal{A}_\mu$ is implied by the vector condition $v_\mu=0$ as long as the zeroth copy scalar $\Phi$ is on-shell.

\section{Generating Charged Solutions}\label{sec:generating}

The classical double copy for Kerr-Schild spacetimes offers a new way of understanding solutions in a sector of general relativity where the Einstein equations linearize, by mapping to an Abelian gauge theory where the equations of motion are always linear. For instance, the Kerr solution is written as the double copy of the so-called $\sqrt{\text{Kerr}}$ gauge field, which provides an explanation for some properties of probe scattering in a Kerr background \cite{Arkani-Hamed:2019ymq}. We can also look at this relationship from the perspective of the single copy theory. The single copy equations of motion are much easier to solve than the Einstein equations, even with the Kerr-Schild ansatz. By combining those equations with the requirement that the gauge field can be obtained from a single copy (that is, in the Kerr-Schild case, the vector $k_\mu$ is null and geodesic), the possible single copy solutions are restricted to a small class \cite{Bah:2019sda}. This helps to explain the Janis-Newman trick for deriving the Kerr solution, and provides some physical understanding of Kerr-Schild solutions.

In this section we take a first look at how these ideas apply to the sKS double copy. In Section \ref{sec:boost_reduction}, we exhibit the boost-reduction procedure introduced in \cite{frolovChargedRotatingBlack1987a} for charging a black hole solution via a higher-dimensional embedding, and show how this amounts to combining a zeroth copy, single copy, and double copy system into a single Einstein-Maxwell-dilaton solution. In Section \ref{sec:consistency}, we give the necessary conditions for a gauge field and scalar to be the single copy of an sKS solution. Finally, in Section \ref{sec:multicharge}, we discuss possibilities for going beyond the sKS ansatz to include solutions with multiple charges, and how we might formulate a single copy in such cases.

\subsection{Boost-Reduction}\label{sec:boost_reduction}

In \cite{frolovChargedRotatingBlack1987a} (see also \cite{Lee:2018gxc}), it is shown how to construct a charged 4D Einstein-Maxwell-dilaton solution from a 5D perspective. One starts from the product of a 4D Kerr spacetime with an $S^1$ and performs a boost along the $S^1$ direction. Reducing on the $S^1$, the result is a black hole with a gauge field and a dilaton, furnishing a charged solution to an action of the form \eqref{eq:emd_action}.

Let us exhibit how this procedure works in light of the sKS and KS double copy relations. For the sake of simplicity, we will start from a 4D Schwarzschild seed, before explaining the procedure for a rotating seed. The Schwarzschild solution in Kerr-Schild coordinates is
\begin{equation}
    g_{\mu\nu} = \eta_{\mu\nu} + \frac{\mu}{r} k_\mu k_\nu,\quad\text{with}\quad k_\mu dx^\mu = dt + dr.
\end{equation}
Its single copy is the Coulomb field
\begin{equation}
    A_\mu\,dx^\mu = \frac{\mu}{r}\left(dt + dr\right),
\end{equation}
and its zeroth copy is $\frac{\mu}{r}$.

Now we take the product with an $S^1$ with coordinate $z$, and boost along the $z$ direction via
\begin{equation}
\begin{split}
    dt \mapsto \cosh\beta\,dt + \sinh\beta\,dz,\\
    dz \mapsto \sinh\beta\,dt + \cosh\beta\,dz.
\end{split}
\end{equation}
The boost leaves the background $\eta_{ab}$ invariant, but modifies the Kerr-Schild vector, so the 5D metric becomes
\begin{equation}\label{eq:schwarzschild_boost}
    g_{ab} = \eta_{ab} + \frac{\mu \sinh^2\beta}{r} k_a k_b,\quad\text{with}\quad k_a dx^a = \frac{\cosh\beta\,dt + dr + \sinh\beta\,dz}{\sinh\beta}.
\end{equation}
Note we moved a factor of $\sinh^2\beta$ into the Kerr-Schild scalar so that we would have $k_5 = 1$, as per the assumption in our reduction procedure in section \ref{sec:reduction}.

After reducing \eqref{eq:schwarzschild_boost} on the $S^1$, we have an sKS solution
\begin{equation}\label{eq:boost_reduce_sks}
    \begin{split}
        g_{\mu\nu} &= e^{-\phi}\left(\eta_{\mu\nu} + \frac{\mu \sinh^2\beta}{r + \mu\sinh^2\beta} k_\mu k_\nu\right),\\
        A_\mu &= \frac{\mu \sinh^2\beta}{r + \mu\sinh^2\beta} k_\mu,\\
        e^{-2\phi} &= 1 + \frac{\mu \sinh^2\beta}{r}
    \end{split}
\end{equation}
with $k_\mu dx^\mu = \frac{\cosh\beta\,dt + dr}{\sinh\beta}$. The sKS single copy is by construction the dimensional reductions of the boosted KS single copy, or
\begin{equation}
    \left(\mathcal{A}_\mu dx^\mu, \Phi\right) = \left(\frac{\mu \sinh\beta}{r}\left(\cosh\beta\,dt + dr\right), \frac{\mu \sinh^2\beta}{r}\right).
\end{equation}

We can compare this to the single copy of the original 4D Schwarzschild seed. Again we have a Coulomb gauge field and a scalar proportional to $r^{-1}$, but they have been rescaled by $\sinh\beta\cosh\beta$ and $\sinh^2\beta$, respectively. 

We can better understand the physical implication of this rescaling by looking at the mass $M$, electric charge $Q$, and scalar charge $\Sigma$ in the sKS solution \eqref{eq:boost_reduce_sks}. These are given by
\begin{equation}\label{eq:adm_charges}
\begin{split}
	M &= \frac{1}{4\pi}\int_S \xi^\mu \left(\nabla_\mu \xi_\nu\right)\,n^\nu\,dA = \frac{\mu}{4}\left(1 + \cosh^2 \beta\right),\\
	Q &= \frac{1}{4\pi}\int_S \xi^\mu F_{\mu\nu} \,n^\nu\,dA = \mu \sinh\beta\cosh\beta,\\
	\Sigma &= \frac{1}{4\pi}\int_S \left(\nabla_\mu \phi\right)n^\mu \,dA = \frac{\mu}{2} \sinh^2\beta.
\end{split}
\end{equation}
where $\xi = \partial_t$ is the timelike Killing vector and $n$ is a unit normal to the integration surface. This family of solutions is considered in \cite{Gibbons:1985ac}, and the extremality bound for coordinate singularities to be shielded by a horizon is found to be\footnote{In \cite{Gibbons:1985ac}, the action and definitions of charges are such that $Q_\text{theirs} = \frac{1}{2}Q_\text{ours}$ and $\Sigma_\text{theirs} = \frac{\sqrt{3}}{8\pi}\Sigma_\text{ours}$.}
\begin{equation}\label{eq:extremal}
	Q^2 \le 4M^2 + 3\Sigma^2.
\end{equation}
We see that this is saturated as $\beta\to\infty$, so the causality bound on the boost in 5D corresponds to the extremality bound of the charged solution in 4D.

This correspondence is well-known, but the double copy gives an alternative perspective on it. Starting from an ordinary instance of the double copy in $D$ dimensions, we can take the double copy spacetime $\mathcal{M}$ and consider the $D+1$-dimensional solution $\mathcal{M}\times S^1$. We can then perform a boost, mixing the two Killing directions in the $(D+1)$-dimensional solution, and compactify the $S^1$ direction. We recover a one-parameter family of sKS solutions that combine the original zeroth copy scalar, single copy gauge field, and double copy metric, with their charges rescaled so as to obey the extremality bound \eqref{eq:extremal}.

With a Schwarzschild seed, it is less surprising that the zeroth, single, and double copies can be combined into a single solution, since spherical symmetry restricts all the fields. It is more striking that the Kerr solution can be combined with its so-called $\sqrt{\text{Kerr}}$ single copy to give a charged and rotating sKS solution. In Boyer-Lindquist coordinates, the Kerr solution is a Kerr-Schild transformation of flat space with
\begin{equation}
	\phi = \frac{\mu r}{r^2 + a^2\cos^2\theta}, \qquad k^\mu = \left(1, 1, 0, -\frac{a}{r^2+a^2}\right).
\end{equation}
The boost-reduction procedure gives the sKS solution
\begin{equation}
\begin{split}
	g_{\mu\nu} &= e^{-\phi}\left(\eta_{\mu\nu} + \frac{\mu r \sinh^2\beta}{r^2 + \mu r \sinh^2\beta + a^2\cos^2\theta} \tilde k_\mu \tilde k_\nu\right),\\
	A_\mu &= \frac{\mu r \sinh^2\beta}{r^2 + \mu r \sinh^2\beta + a^2\cos^2\theta} \tilde k_\mu,\\
	e^{-2\phi} &= 1 + \frac{\mu r \sinh^2\beta}{r^2 + \mu r \sinh^2\beta + a^2\cos^2\theta},
\end{split}
\end{equation}
where $\tilde k^\mu = \frac{1}{\sinh\beta} \left(\cosh\beta, 1, 0, -\frac{a}{r^2+a^2}\right)$. This combines the Kerr metric with its single copy gauge field and zeroth copy scalar into a single sKS solution.

The angular momentum of this solution is given by
\begin{equation}
	J = \frac{1}{4\pi}\int_S \xi^\mu \left(\nabla_\mu \zeta_\nu\right)n^\nu\,dA = a\mu \cosh\beta,
\end{equation}
where $\zeta = \partial_\phi$ is the azimuthal Killing vector and $\xi$ is a timelike Killing vector with $\xi^\mu \zeta_\mu = 0$. The charges $M$, $Q$, and $\Sigma$ are independent of $a$ and are the same as in \eqref{eq:adm_charges}. As $\beta\to\infty$ we have $J/Q \to 0$, so the extremal bound \eqref{eq:extremal} is unaffected.

\subsection{Reconstructing sKS Solutions}\label{sec:consistency}

The core simplification underlying the classical Kerr-Schild double copy is the fact that for a Kerr-Schild metric, the Ricci tensor with mixed indices becomes linear in the scalar $\Phi$ \cite{Monteiro:2014cda}. This is the reason we can draw a precise relationship with a solution to the linear Maxwell equations. The double copy thus gives a new perspective on the long-understood fact that Kerr-Schild metrics represent a more tractable sector of general relativity. By solving the single copy equations of motion, we can hope to reconstruct exact double copy metrics from them.

This procedure will never be guaranteed to reproduce double copy solutions, because there are more equations of motion for the double copy theory than for the single copy theory. Solving the single copy equations of motion is a necessary but not sufficient condition for the associated double copy fields to form an exact solution. If we try to impose the additional equations of motion from the double copy theory, the resulting problem is no easier than if we were to seek double copy solutions directly.

For the Kerr-Schild double copy, there is a middle ground that avoids these difficulties, but still imposes enough of a constraint on single-copy solutions that we often end up being able to reconstruct bona fide double copies. Since the Kerr-Schild metrics are formed from a null geodesic vector $k_\mu$, the single-copy gauge field $A_\mu = \Phi k_\mu$ must also be null and geodesic in an appropriate gauge. As discussed in \cite{Carrillo-Gonzalez:2017iyj,Bah:2019sda}, this implies that the gauge field is an eigenvector of its own field strength,
\begin{equation}\label{eq:eigenvector}
	{F^\mu}_\nu A^\nu = \lambda A^\nu,
\end{equation}
for some scalar function $\lambda$. In \cite{Bah:2019sda}, it is shown that this leads to a necessary condition for single copy gauge fields in terms of classical scattering, and that in four dimensions the gauge fields satisfying \eqref{eq:eigenvector} correspond to Li\'enard-Wiechert fields with a possible complex shift. The simplest examples of these Li\'enard-Wiechert fields all generate exact double copy solutions, even though \emph{a priori} the double copy equations of motion were not guaranteed to be satisfied.

Here we discuss how this approach may be extended to the sKS double copy. Given a solution to the single copy theory, i.e. a scalar field $\Phi$ solving the Klein-Gordon equation and gauge vector $\mathcal{A}_\mu$ solving Maxwell's equations, we would like to reconstruct a solution to the sKS equations of motion,
\begin{equation}
(\mathcal{A}_\mu, \Phi) \quad \xrightarrow{\text{Double Copy}} \quad (g^{(D)}_{\mu \nu}, A_\mu, \phi).
\end{equation}
Equations \eqref{eq:sks_reduced} and \eqref{eq:sks_sc} can be inverted to give a candidate sKS solution, 
\begin{align}\label{reconstructed_sKS}
g_{\mu \nu}^{(D)}&=(\Phi+1)^{\frac{1}{D-2}} \left[ \bar{g}_{\mu \nu}+\Phi^{-1} (\Phi+1)^{-1} \mathcal{A}_\mu \mathcal{A}_\nu\right]
\\
\phi&=-\frac{\log(\Phi+1)}{D-2}, \quad  A_\mu =(\Phi+1)^{-1} \mathcal{A}_\mu. \nonumber
\end{align}
To ensure that the timelike sKS vector $k_\mu=\Phi^{-1} \mathcal{A}_\mu$ is appropriately normalized, with $\bar{g}^{\mu \nu} k_\mu k_\nu=-1$, the sKS single and zeroth copy fields must be related by
\begin{equation}\label{gauge_condition}
\mathcal{A}_\mu \mathcal{A}^\mu =-\Phi^2.
\end{equation}
Similarly, the requirement that $k_\mu$ be a geodesic vector field enforces
\begin{equation}\label{geodesic_condition}
\mathcal{A}_\nu \bar{\nabla}^\nu \mathcal{A}_\mu= \chi \mathcal{A}_\mu
\end{equation}
for some scalar function $\chi$. For example, $k_\mu$ will be affinely parameterized if and only if $\chi = k_\mu \partial^\mu \Phi$. Together the conditions \eqref{gauge_condition} and \eqref{geodesic_condition} restrict the class of gauge fields admitting canonically normalized sKS double copies. 

If we try to form a condition analogous to \eqref{eq:eigenvector} directly for the sKS single copy, we find a modification due to the fact that the gauge field is no longer null. The gauge field is not an eigenvector of its field strength, but rather satisfies
\begin{equation}\label{eq:modified_eigenvector}
	{\mathcal{F}^\mu}_\nu \mathcal{A}^\nu = -\chi \mathcal{A}^\mu - \Phi \bar{\nabla}^\mu \Phi,
\end{equation}
where ${\mathcal{F}^\mu}_\nu = \bar{\nabla}^\mu \mathcal{A}_\nu - \bar{\nabla}_\nu \mathcal{A}^\mu$.

It would be interesting to understand the solutions to \eqref{eq:modified_eigenvector} in general, but for the moment we restrict our focus to one way in which a solution to \eqref{eq:modified_eigenvector} can arise from a solution to \eqref{eq:eigenvector} on a flat background. Let $A^\mu$ be a time-independent solution to \eqref{eq:eigenvector}, with $A^\mu = \Phi k^\mu$ in a normalization where $k^0 = 1$ and with $k^\mu$ an affinely parametrized geodesic, as in the original Kerr-Schild double copy of \cite{Monteiro:2014cda}. We can then take
\begin{equation}
	\mathcal{A}^0 = A^0 + \alpha \Phi, \qquad \mathcal{A}^i = A^i,
\end{equation}
where the background metric is in coordinates where $\eta_{00} = -1$ and the index $i$ runs from 1 to $D-1$. The field strength is given by
\begin{equation}
	{\mathcal{F}^\mu}_0 = {F^\mu}_0 - \alpha \bar{\nabla}^\mu \Phi, \qquad {\mathcal{F}^0}_\nu = {F^0}_\nu - \alpha \bar{\nabla}_\nu \Phi.
\end{equation}
It follows that
\begin{equation}
\begin{split}
	{\mathcal{F}^\mu}_\nu \mathcal{A}^\nu &= -\chi \mathcal{A}^\mu - (2\alpha + \alpha^2) \Phi \bar{\nabla}^\mu \Phi,
\end{split}
\end{equation}
where we used the condition $\chi = k_\mu \bar{\nabla}^\mu \Phi$ for $k$ to be affinely parametrized, and applied \eqref{eq:eigenvector}. We see that $\frac{1}{\sqrt{(\alpha + 1)^2 - 1}}\mathcal{A}^\mu$ will satisfy \eqref{eq:modified_eigenvector} with the scalar $\Phi$ and a rescaled $\chi$. Comparing with the previous section, we see that we should identify $\alpha = \cosh \beta - 1$, in which case the prefactor is $\frac{1}{\sinh\beta}$. In summary, given a solution $A^\mu = \Phi k^\mu$ to \eqref{eq:eigenvector} with $k$ an affinely parametrized null geodesic and $k^0 = 1$, we can form a solution
\begin{equation}
	\mathcal{A}^\mu = \frac{\Phi}{\sinh\beta}\left(\cosh \beta, k^i\right)
\end{equation}
to \eqref{eq:modified_eigenvector}. These solutions are the single copies of the sKS solutions formed via boost-reduction as in Section \ref{sec:boost_reduction}, and following the map \eqref{reconstructed_sKS} we can form the associated sKS solutions from them.

Another approach to \eqref{eq:modified_eigenvector} is to directly employ the fact that, together with \eqref{gauge_condition}, it implies \eqref{eq:eigenvector} for an uplift gauge field $\mathcal{A}^a$ formed according to \eqref{eq:vector_reduction}. That is, given any candidate sKS single copy gauge field and scalar, we can form a $(D+1)$-dimensional gauge field from them and check whether it satisfies \eqref{eq:eigenvector}. This is a necessary condition for the $D$-dimensional gauge field and scalar to be the single copy of an sKS solution.

\subsection{Beyond sKS: Multiple Charges}\label{sec:multicharge}

Ans\"atze similar to sKS have been explored for multi-charge solutions, e.g. in \cite{Wu:2011gp,Wu:2011gq}. However, since these more general ans\"atze do not come from a KK reduction of an ordinary Kerr-Schild spacetime, the method taken in this paper for exhibiting a single copy of such solutions does not directly apply. Nevertheless, there is reason to believe that the double copy also applies to some of these multi-charge solutions. Here we consider a simple example, in which the the metric and gauge fields map to a set of several sKS single copies\footnote{We thank Pierre Heidmann for suggesting this notion of the single copies of a multi-charge solution.}.

Consider the three-charge $AdS_5$ black hole of $\mathcal{N}=2$ gauged $U(1)^3$ supergravity, which can be obtained in a consistent truncation of type IIB supergravity on $S^5$ \cite{Behrndt:1998jd,cvetic:1999}:
\begin{equation}\label{eq:AdS5_BH_metric}
    ds^2 = -\lambda^{-2/3}f\,dt^2 + \lambda^{1/3}\left(f^{-1} dr^2 + r^2 \,d\Omega_3^2\right), \qquad A^{(i)}=(1-H_i^{-1}) \coth \beta_i \, dt
\end{equation}
where $i=1,2,3$, and 
\begin{equation}
\lambda=\prod_{i=1}^3 H_i, \qquad H_i=1+\frac{\mu \sinh^2 \beta_i}{r^2}, \qquad f=1-\frac{\mu}{r^2}+g^2 r^2 \lambda.
\end{equation}
Solutions of this form can be viewed as one-parameter, non-extremal deformations of corresponding BPS-saturated solutions, with $\mu$ parameterizing the non-extremality \cite{cvetic:1996}. 
While \eqref{eq:AdS5_BH_metric} is not strictly speaking sKS, owing to the multiple gauge vectors, it admits what we might call a multi-sKS form, like the ansatz introduced in \cite{Wu:2011gq}. 
Under the change of coordinates
\begin{equation} \label{multi_sKS_coord}
dt=d\bar{t}-\sqrt{\left(\frac{\mu}{r^2}-g^2r^2\lambda+g^2 r^2\right)\left(\frac{\mu}{r^2}+\lambda-1\right)} \,\, \frac{ dr}{f(1+g^2 r^2)},
\end{equation}
the metric takes the form
\begin{align}\label{eq:multi_sKS}
ds^2=\lambda^{\frac{1}{3}}\left[ -(1+g^2r^2)d\bar{t}^2+\frac{dr^2}{1+g^2 r^2} +r^2 d\Omega_3^2+\sum_{i=1}^3(1-H_i^{-1})(k_\mu^{(i)} dx^\mu)^2 \right], \nonumber
\\
k_\mu^{(i)} dx^\mu=\frac{\cosh \delta_i \sinh \delta_i }{\prod_{j \neq i} \sqrt{\sinh^2 \delta_i-\sinh^2 \delta_j}}\left(d\bar{t}+ \sqrt{\frac{\frac{\mu}{r^2}-g^2 r^2(\lambda-1)}{\frac{\mu}{r^2}+\lambda-1}}  \,\, \frac{dr}{1+g^2 r^2}\right).
\end{align}
In any given single-charge limit, e.g. $\beta_1 = \beta$ with $\beta_2, \beta_3 \rightarrow 0$, the functions $H_2$, $H_3$ go to unity so that the terms proportional to $1-H_2^{-1}$ and $1-H_3^{-1}$ vanish, and we are left with a single-sKS metric \eqref{eq:sks_reduced} with $e^{-(D-2)\phi}=H_1$, $k_\mu=k_\mu^{(1)}$, and $A=A^{(1)}$. For each such limit we can apply the double copy relations \eqref{eq:sks_sc} to extract a well-defined single copy. In effect we have three sKS single copies,
\begin{equation}\label{multi-sKS_single_copy}
\mathcal{A}_\mu^{(i)} \equiv H_i A_\mu^{(i)}, \qquad \Phi_i \equiv H_i-1,
\end{equation}
each guaranteed to satisfy the Maxwell-scalar equations of motion on the background spacetime. Crucially, the gauge vectors $A^{(i)}$ and harmonic functions $H_i$ are independent of the $j \neq i$ charge parameters, so that these $(\mathcal{A}_\mu^{(i)},\Phi_i)$ can be regarded as the single copy of the full three-charge solution.  In the present case,
\begin{equation}
\mathcal{A}_\mu^{(i)}dx^\mu =\frac{\mu \cosh \beta_i \sinh \beta_i}{r^2} \, d\bar{t}, \qquad \Phi_i=\frac{\mu \sinh^2 \beta_i}{r^2},
\end{equation}
up to radially dependent gauge transformation of the vectors. The full solution \eqref{eq:multi_sKS} can be thought of as the double copy of these three pairs of gauge fields and scalars. Similarly, the four-charge AdS$_4$ and two-charge AdS$_2$ black hole solutions of \cite{cvetic:1999} can be mapped to sets of four and two sKS single copies, respectively. Given that the multi-sKS structure of \eqref{eq:AdS5_BH_metric} was explicitly used to construct its doubly rotating cousin \cite{Wu:2011gq}, we anticipate that an analogous application of the sKS double copy can be employed there as well.

These examples clearly show that there are solutions more general than the sKS ansatz, incorporating multiple charges, for which a multi-sKS single copy applies. However, the most general statement of a multi-sKS ansatz and associated single copy is not yet clear.
In the multi-sKS form \eqref{eq:multi_sKS} for the static three-charge AdS$_5$ black hole, the vectors $k^{(i)}$ do not obey a fixed, time-like normalization except in the appropriate single-charge limit, and their relationship to the gauge fields $A^{(i)}$ is modified away from the single sKS expectation $(1-H_i^{-1})k^{(i)}$ by constant factors depending on the charge parameters $\beta_i$. Even so, the dependence of the $A^{(i)}$ and $H_i$ on a single charge parameter $\beta_i$ allows for the sKS single copies of the single-charge limits to be interpreted together as the single copy of the full solution.
 It will be interesting to understand how far the sKS double copy can be extended in this direction to include the many multi-charge solutions arising in consistent truncations of supergravity theories.

\section{Conclusion}

We have shown that under compactification on a circle, the Kerr-Schild double copy of \cite{Monteiro:2014cda} descends to a notion of the double copy for Kaluza-Klein gravity. Given a solution of a Kaluza-Klein theory admitting a presentation in stringy Kerr-Schild (sKS) form \cite{Wu:2011zzh,Wu:2015nra}, we can write it as the double copy of a gauge field and a scalar, which are not coupled in the single copy theory. Analogous to how the ordinary Kerr-Schild double copy expresses a spin-2 field as the square of a spin-1 field, we see that the spin-0, spin-1, and spin-2 fields comprising an sKS solution can be written as the square of the combination of spin-0 and spin-1 fields in the single copy theory. This is similar to the double copy prescription derived previously for ungauged half-maximal supergravity solutions using double field theory \cite{Angus:2021zhy}. We find furthermore that our double copy prescription holds for an sKS solution to a gauged supergravity theory on a maximally symmetric background spacetime, provided that the zeroth copy scalar is given a mass proportional to the background curvature, in analogy with the results of \cite{Carrillo-Gonzalez:2017iyj}.

Although the sKS double copy stands on its own as a relation between exact classical solutions, it is illuminating to consider its uplift to one higher dimension, where we recover the ordinary Kerr-Schild double copy. We show that, following the approach of \cite{frolovChargedRotatingBlack1987a}, we can start from any instance of the Kerr-Schild double copy for a stationary spacetime, take the product with a compact dimension, boost along the compact dimension, and finally reduce along the same compact dimension. This boost-reduction procedure is well-known to generate charged black hole solutions \cite{frolovChargedRotatingBlack1987a}, but we see furthermore that it explicitly mixes the zeroth, single, and double copy fields of the Kerr-Schild seed into a single sKS solution. Many instances of the sKS double copy can be constructed in this manner.

There is a substantial body of work on the classical double copy, and many of the established results lead us to pose analogous questions for the sKS double copy. Perhaps the most fundamental of these is the alternative formulation of the classical double copy in which the Weyl tensor of the gravity solution is written in terms of the field strength of the gauge field \cite{Luna:2018dpt}. This prescription, known as the Weyl double copy, applies to any four-dimensional spacetime of Petrov type D, including some metrics which do not have a Kerr-Schild form. Furthermore, there is a more general understanding of the relationship that applies to spacetimes of type D in higher dimensions \cite{Alawadhi:2019urr,Alawadhi:2020jrv,Chacon:2021wbr}. It would be very interesting to understand how the Weyl double copy behaves under a compactification, and to see how it might extend the domain of applicability of the double copy for Kaluza-Klein gravity beyond the sKS ansatz.

It would also be interesting to study whether known generalizations of the sKS ansatz admit a double copy interpretation. For instance, in \cite{Wu:2011zzh}, it is shown that there is a natural way to turn on NUT charges for rotating, single-charge AdS black holes of sKS form. The resulting ansatz can be written as the compactification of a multi-Kerr-Schild metric in $D+1$ dimensions. Similarly, the Taub-NUT spacetime in four dimensions can be written in double Kerr-Schild form, and in $D$ dimensions it can analogously be written in a multi-Kerr-Schild form with $\lfloor D/2\rfloor$ mutually orthogonal null geodesics \cite{Chen:2007fs}. This multi-Kerr-Schild coordinate system enables a double copy interpretation of these metrics in four and higher dimensions \cite{Luna:2015paa,Chawla:2022ogv}. The connection between these results and the generalization of the sKS ansatz to include NUT charge is worthy of further study.

In a similar vein, we saw in Section \ref{sec:multicharge} how the double copy might be extended to multi-charged solutions that cannot be expressed as a compactification of a Kerr-Schild spacetime along an $S^1$. This result indicates that, although the Kerr-Schild double copy is useful as a mechanism for deriving the sKS double copy, the latter may apply more broadly than just in cases arising from a compactification of a Kerr-Schild solution. We leave further exploration of the single copies of multi-charge solutions to future work.

\section*{Acknowledgments}

We thank Ibrahima Bah and Pierre Heidmann for useful conversations and feedback regarding this work. The work of PW is supported in part by NSF grant PHY-2112699, and RD is supported by an NSF Graduate Research Fellowship.

\bibliographystyle{unsrt}
\bibliography{main}

\begin{thebibliography}{10}

\bibitem{Kawai:1985xq}
H. Kawai, D.~C. Lewellen, and S.~H.~H. Tye.
\newblock A relation between tree amplitudes of closed and open strings.
\newblock {\em Nuclear Physics B}, 269(1):1--23, May 1986.

\bibitem{Bern:2008qj}
Z. Bern, J.~J.~M. Carrasco, and H. Johansson.
\newblock New {{Relations}} for {{Gauge-Theory Amplitudes}}.
\newblock {\em Phys. Rev. D}, 78:085011, 2008.

\bibitem{Bern:2019prr}
Z. Bern, J.~J. Carrasco, M. Chiodaroli, H. Johansson, and R. Roiban.
\newblock The {{Duality Between Color}} and {{Kinematics}} and its
  {{Applications}}, September 2019.

\bibitem{adamoSnowmassWhitePaper2022}
T. Adamo, J.~J.~M. Carrasco, M. {Carrillo-Gonz{\'a}lez}, M. Chiodaroli, H.
  Elvang, H. Johansson, D. O'Connell, R. Roiban, and O. Schlotterer.
\newblock Snowmass {{White Paper}}: The {{Double Copy}} and its
  {{Applications}}.
\newblock In {\em 2022 {{Snowmass Summer Study}}}, April 2022.

\bibitem{Monteiro:2014cda}
R. Monteiro, D. O'Connell, and C.~D. White.
\newblock Black holes and the double copy.
\newblock {\em Journal of High Energy Physics}, 2014(12):56, December 2014.

\bibitem{Monteiro:2020plf}
R. Monteiro, D. O'Connell, D. Peinador~Veiga, and M. Sergola.
\newblock Classical solutions and their double copy in split signature.
\newblock {\em JHEP}, 05:268, 2021.

\bibitem{Chacon:2021hfe}
E. Chac{\'o}n, A. Luna, and C.~D. White.
\newblock Double copy of the multipole expansion.
\newblock {\em Phys. Rev. D}, 106(8):086020, 2022.

\bibitem{Cho:2019ype}
W. Cho and K. Lee.
\newblock Heterotic {{Kerr-Schild Double Field Theory}} and {{Classical Double
  Copy}}.
\newblock {\em Journal of High Energy Physics}, 2019(7):30, July 2019.

\bibitem{Bahjat-Abbas:2020cyb}
N. {Bahjat-Abbas}, R. {Stark-Much{\~a}o}, and C.~D. White.
\newblock Monopoles, shockwaves and the classical double copy.
\newblock {\em JHEP}, 04:102, 2020.

\bibitem{Berman:2020xvs}
D.~S. Berman, K. Kim, and K. Lee.
\newblock The {{Classical Double Copy}} for {{M-theory}} from a {{Kerr-Schild
  Ansatz}} for {{Exceptional Field Theory}}.
\newblock {\em Journal of High Energy Physics}, 2021(4):71, April 2021.

\bibitem{Luna:2015paa}
A. Luna, R. Monteiro, D. O'Connell, and C.~D. White.
\newblock The classical double copy for {{Taub}}\textendash{{NUT}} spacetime.
\newblock {\em Phys. Lett. B}, 750:272--277, 2015.

\bibitem{Carrillo-Gonzalez:2017iyj}
M. {Carrillo-Gonz{\'a}lez}, R. Penco, and M. Trodden.
\newblock The classical double copy in maximally symmetric spacetimes.
\newblock {\em JHEP}, 04:028, 2018.

\bibitem{CarrilloGonzalez:2019gof}
M. Carrillo~Gonz{\'a}lez, B. Melcher, K. Ratliff, S. Watson, and C.~D. White.
\newblock The classical double copy in three spacetime dimensions.
\newblock {\em JHEP}, 07:167, 2019.

\bibitem{Luna:2016due}
A. Luna, R. Monteiro, I. Nicholson, D. O'Connell, and C.~D. White.
\newblock The double copy: {{Bremsstrahlung}} and accelerating black holes.
\newblock {\em JHEP}, 06:023, 2016.

\bibitem{Bahjat-Abbas:2017htu}
N. {Bahjat-Abbas}, A. Luna, and C.~D. White.
\newblock The {{Kerr-Schild}} double copy in curved spacetime.
\newblock {\em Journal of High Energy Physics}, 2017(12):4, December 2017.

\bibitem{Alkac:2021bav}
G. Alkac, M.~K. Gumus, and M. Tek.
\newblock The {{Kerr-Schild Double Copy}} in {{Lifshitz Spacetime}}.
\newblock {\em JHEP}, 05:214, 2021.

\bibitem{Berman:2018hwd}
D.~S. Berman, E. Chac{\'o}n, A. Luna, and C.~D. White.
\newblock The self-dual classical double copy, and the {{Eguchi-Hanson}}
  instanton.
\newblock {\em JHEP}, 01(arXiv:1809.04063):107, 2019.

\bibitem{Easson:2022zoh}
D.~A. Easson, T. Manton, and A. Svesko.
\newblock {Einstein-Maxwell theory and the Weyl double copy}.
\newblock 10 2022.

\bibitem{Angus:2021zhy}
S. Angus, K. Cho, and K. Lee.
\newblock {The classical double copy for half-maximal supergravities and
  T-duality}.
\newblock {\em JHEP}, 10:211, 2021.

\bibitem{Luna:2018dpt}
A. Luna, R. Monteiro, I. Nicholson, and D. O'Connell.
\newblock Type {{D Spacetimes}} and the {{Weyl Double Copy}}.
\newblock {\em Classical and Quantum Gravity}, 36(6):065003, March 2019.

\bibitem{White:2020sfn}
C.~D. White.
\newblock A {{Twistorial Foundation}} for the {{Classical Double Copy}}.
\newblock {\em Physical Review Letters}, 126(6):061602, February 2021.

\bibitem{Luna:2022dxo}
A. Luna, N. Moynihan, and C.~D. White.
\newblock Why is the {{Weyl}} double copy local in position space?, August
  2022.

\bibitem{Wu:2011zzh}
S.-Q. Wu.
\newblock General rotating charged {{Kaluza-Klein AdS}} black holes in higher
  dimensions.
\newblock {\em Physical Review D}, 83(12):121502, June 2011.

\bibitem{Wu:2015nra}
S.-Q. Wu and H. Wang.
\newblock Approach of background metric expansion to a new metric ansatz for
  gauged and ungauged {{Kaluza-Klein}} supergravity black holes.
\newblock {\em Physical Review D}, 91(10):104031, May 2015.

\bibitem{frolovChargedRotatingBlack1987a}
V.~P. Frolov, A.~I. Zelnikov, and U. Bleyer.
\newblock Charged {{Rotating Black Hole From Five-dimensional Point}} of
  {{View}}.
\newblock {\em Annalen Phys.}, 44:371--377, 1987.

\bibitem{Ett:2015fhw}
B. Ett.
\newblock Exact {{Solutions}} in {{Gravity}}: {{A}} journey through spacetime
  with the {{Kerr-Schild}} ansatz.
\newblock {\em Doctoral Dissertations}, November 2015.

\bibitem{Banerjee:2019saj}
A. Banerjee, E.~O. Colg\'ain, J.~A. Rosabal, and H. Yavartanoo.
\newblock {Ehlers as EM duality in the double copy}.
\newblock {\em Phys. Rev. D}, 102:126017, 2020.

\bibitem{Mandal:2000rp}
G. Mandal.
\newblock A review of the {{D1}}/{{D5}} system and five dimensional black hole
  from supergravity and brane viewpoint, March 2000.

\bibitem{callan:1996}
C.~G. Callan and J.~M. Maldacena.
\newblock D-brane approach to black hole quantum mechanics.
\newblock {\em Nuclear Physics B}, 472(3):591--608, 1996.

\bibitem{Gibbons:2004}
G.~W. Gibbons, H. Lü, D.~N. Page, and C.~N. Pope.
\newblock Rotating black holes in higher dimensions with a cosmological
  constant.
\newblock {\em Phys. Rev. Lett.}, 93:171102, Oct 2004.

\bibitem{cvetic:1999}
M. Cvetič, M. Duff, P. Hoxha, J.~T. Liu, H. Lü, J. Lu, R. Martinez-Acosta, C.
  Pope, H. Sati, and T. Tran.
\newblock Embedding {AdS} black holes in ten and eleven dimensions.
\newblock {\em Nuclear Physics B}, 558(1-2):96--126, 1999.

\bibitem{Bah:2019sda}
I. Bah, R. Dempsey, and P. Weck.
\newblock Kerr-{{Schild Double Copy}} and {{Complex Worldlines}}.
\newblock {\em Journal of High Energy Physics}, 2020(2):180, February 2020.

\bibitem{Arkani-Hamed:2019ymq}
N. {Arkani-Hamed}, Y.-t. Huang, and D. O'Connell.
\newblock Kerr {{Black Holes}} as {{Elementary Particles}}.
\newblock {\em Journal of High Energy Physics}, 2020(1):46, January 2020.

\bibitem{Lee:2018gxc}
K. Lee.
\newblock {Kerr-Schild Double Field Theory and Classical Double Copy}.
\newblock {\em JHEP}, 10:027, 2018.

\bibitem{Gibbons:1985ac}
G.~W. Gibbons and D.~L. Wiltshire.
\newblock Black holes in {{Kaluza-Klein}} theory.
\newblock {\em Annals of Physics}, 167(1):201--223, March 1986.

\bibitem{Wu:2011gp}
S.-Q. Wu.
\newblock {Two-charged non-extremal rotating black holes in seven-dimensional
  gauged supergravity: The Single-rotation case}.
\newblock {\em Phys. Lett. B}, 705:383--387, 2011.

\bibitem{Wu:2011gq}
S.-Q. Wu.
\newblock {General Nonextremal Rotating Charged AdS Black Holes in
  Five-dimensional $U(1)^3$ Gauged Supergravity: A Simple Construction Method}.
\newblock {\em Phys. Lett. B}, 707:286--291, 2012.

\bibitem{Behrndt:1998jd}
K. Behrndt, M. Cvetic, and W.~A. Sabra.
\newblock {Nonextreme black holes of five-dimensional N=2 AdS supergravity}.
\newblock {\em Nucl. Phys. B}, 553:317--332, 1999.

\bibitem{cvetic:1996}
M. Cvetič and A. Tseytlin.
\newblock Non-extreme black holes from non-extreme intersecting m-branes.
\newblock {\em Nuclear Physics B}, 478(1):181--198, 1996.

\bibitem{Alawadhi:2019urr}
R. Alawadhi, D.~S. Berman, B. Spence, and D.~P. Veiga.
\newblock S-duality and the {{Double Copy}}.
\newblock {\em Journal of High Energy Physics}, 2020(3):59, March 2020.

\bibitem{Alawadhi:2020jrv}
R. Alawadhi, D.~S. Berman, and B. Spence.
\newblock Weyl doubling.
\newblock {\em Journal of High Energy Physics}, 2020(9):127, September 2020.

\bibitem{Chacon:2021wbr}
E. Chac{\'o}n, S. Nagy, and C.~D. White.
\newblock The {{Weyl}} double copy from twistor space.
\newblock {\em Journal of High Energy Physics}, 2021(5):2239, May 2021.

\bibitem{Chen:2007fs}
W. Chen and H. Lu.
\newblock {Kerr-Schild structure and harmonic 2-forms on (A)dS-Kerr-NUT
  metrics}.
\newblock {\em Phys. Lett. B}, 658:158--163, 2008.

\bibitem{Chawla:2022ogv}
S. Chawla and C. Keeler.
\newblock {Aligned Fields Double Copy to Kerr-NUT-(A)dS}.
\newblock 9 2022.

\end{thebibliography}

\end{document}